\pdfoutput=1 % only if pdf/png/jpg images are used
\documentclass[11pt,a4paper]{article}
\usepackage{graphicx}
\usepackage{jinstpub}
\usepackage[separate-uncertainty,retain-explicit-plus,per-mode=symbol,binary-units]{siunitx}
\usepackage{array,mathtools,amssymb,dcolumn}
\usepackage{rotating}
\usepackage[version=4]{mhchem}
\DeclareSIUnit\year{y}
\DeclareSIUnit\inch{in}
\DeclareSIUnit\c{\mbox{$c$}}
\DeclareSIUnit\electron{\mbox{$e^-$}}
\newcommand{\mbb}{\mbox{$m_{\beta\beta}$}}
\newcommand{\qbb}{\mbox{$Q_{\beta\beta}$}}

\newcommand{\tth}{\mbox{T$^{2\nu}_{1/2}$}}
\title{A high-resolution CMOS imaging detector for the search of neutrinoless double $\beta$ decay in \ce{^{82}Se}}
\author[a,1]{A.E.~Chavarria,
\note{Corresponding author.}}
\author[b]{C.~Galbiati, }
\author[b]{X.~Li }
\author[c]{and J.A.~Rowlands}

\affiliation[a]{Kavli Institute for Cosmological Physics and The Enrico Fermi Institute, The University of Chicago, Chicago, IL, United States}
\affiliation[b]{Department of Physics, Princeton University, Princeton, NJ, United States}
\affiliation[c]{Division of Physical Sciences, Sunnybrook Health Science Centre, University of Toronto, Toronto, ON, Canada}

\emailAdd{alvaro@kicp.uchicago.edu}
\abstract{We introduce high-resolution solid-state imaging detectors for the search of neutrinoless double $\beta$ decay.  Based on the present literature, imaging devices from amorphous \ce{^{82}Se} evaporated on a complementary metal-oxide-semiconductor (CMOS) active pixel array could have the energy and spatial resolution to produce two-dimensional images of ionizing tracks of utmost quality, effectively akin to an electronic bubble chamber in the double $\beta$ decay energy regime.  Still to be experimentally demonstrated, a detector consisting of a large array of these devices could have very low backgrounds, possibly reaching \SI{1E-7}{\per\kg\per\year} in the neutrinoless decay region of interest (ROI), as it may be required for the full exploration of the neutrinoless double $\beta$ decay parameter space in the most unfavorable condition of a strongly quenched nucleon axial coupling constant.}
\notoc

\begin{document}
\maketitle
\flushbottom
%---

\section{Introduction}
The study of double $\beta$ decay is a unique method of inquiry into the nature of neutrinos, as the observation of a neutrinoless branching ratio is the only known avenue to establish that neutrinos are Majorana fermions as opposed to Dirac fermions~\cite{Dirac:1928dh, Dirac:1928cz, Majorana:1937vz,GoeppertMayer:1935ip,Furry:1939ey,Racah:1937dc}.

Recent results on the nuclear physics of the decay~\cite{Barea:2013eu} highlighted the very elusive nature of this transition: for the case of a normal hierarchy of neutrino masses and a strongly quenched nucleon axial coupling constant,\footnote{The strongly quenched nucleon axial coupling constant is phenomenologically parameterized in Ref.~\cite{Barea:2013eu} as $g_{\rm phen.}=1.269\times A^{-0.18}$, where $A$ is the mass number.} a complete program for the discovery of neutrinoless double $\beta$ decay, reaching the Majorana neutrino mass (\mbb) of \SI{1}{\meV\per\square\c}, requires the collection of an exposure of O(\SI{1E7}{\kilo\gram\year}) in a background-free mode.

Following the notation of Ref.~\cite{DellOro:2016gf}, the background-free condition can be expressed as:
\begin{equation}
M \cdot T \cdot B \cdot \Delta \leq 1,
\end{equation}
where $M$ is the mass of isotope, $T$ is the measurement time, $B$ is the background rate per unit mass, energy and time about the $Q$-value of the decay (\qbb ), and $\Delta$ is the full width at half maximum (FWHM) of the energy response at \qbb.  Considering the required exposure (i.e., $M \cdot T$) cited above, the background-free condition is equivalent to the requirement:
\begin{equation}
B \cdot \Delta \leq \SI{1E-7}{\per\kg\per\year}.
\label{eq:BackgroundRequirement}
\end{equation}

Values of $B \cdot \Delta$ for currently operating and future proposed experiments are at least four orders of magnitude higher than this value, see Table~7 in Ref.~\cite{DellOro:2016gf} for a review.

We present a study on the low-background performance of low-noise imaging devices from amorphous \ce{^{82}Se} evaporated on a complementary metal-oxide-semiconductor (CMOS) active pixel array.  Such devices could have the capability to image with high spatial and energy resolution the \si{\mm}-long tracks produced by minimum ionizing electrons emitted in the double $\beta$ decay of \ce{^{82}Se}.  We present estimates of the expected backgrounds for a large imaging detector deployed in a low-radioactivity environment and show that the proposed technology combines at once the possibility of a low-background implementation, the precise energy resolution required to reject background from the two-neutrino double $\beta$ decay channel and the efficient determination of the event topology necessary for a powerful rejection of residual $\gamma$-ray background. While not yet experimentally proven, the proposed detector embodies elements that could allow to achieve the background-free requirement set by Eq.~\ref{eq:BackgroundRequirement}.

We note that the strategies for background suppression presented in this paper apply generally to high-resolution solid-state imaging detectors. We have focused on amorphous selenium technology because it shows the greatest potential at present. However, recent advances in CdTe technology for X-ray imaging~\cite{Bellazzini:2013} could motivate an alternate realization of the proposed concept for the search of neutrinoless double $\beta$ decay in \ce{^{116}Cd}.

\section{Conceptual design}
The isotope \ce{^{82}Se} is a good candidate for the search of neutrinoless double $\beta$ decay~\cite{Arnold:1998ha,Arnold:2006bk,Artusa:2016ek} due to its high $\qbb = \SI{2998}{\keV}$~\cite{Lincoln:2013kt} and relatively long two-neutrino double $\beta$ decay half-life of $\tth = \SI{1E20}{\year}$~\cite{Arnold:2006bk}.  The high \qbb\ leads to a favorable phase-space factor and is significantly above most interfering $\gamma$-ray lines from the uranium and thorium decay chains of natural radioactivity, while the long \tth\ leads to a small contribution of two-neutrino double electron events near \qbb.  In addition, selenium can form a wide variety of gases that allow for isotopic enrichment by centrifugation~\cite{Beeman:2015du} or, potentially, at larger scales by distillation~\cite{Mills:1988tx}.

Amorphous selenium is a well-established photoconductor commonly used as an X-ray converter in large area (\SI{\sim1000}{\square\centi\meter}) flat panel detectors employed in the medical industry for radiographic and fluoroscopic imaging~\cite{Zhao:1995fx,Antonuk:2000fo,Hunt:2004js}.  These detectors feature a thin-film transistor (TFT) pixel array on which a layer of amorphous selenium is evaporated.  In turn, a thin metal electrode is deposited on the back of the device to apply an electric field across the selenium.  Charge carriers produced by ionizing radiation in the selenium are drifted toward the pixel array, where the charge is stored by a small capacitor on the pixel.  Current TFT technology is limited by the relatively large pixel readout noise of $\sim$\SI{1000}{\electron} (charge carriers)~\cite{Antonuk:2000fo}.

The noise can be readily improved by replacing the TFT array with a CMOS active pixel array. Amorphous selenium photoconductive layers have already been successfully interfaced with CMOS pixel arrays in the development of high-spatial-resolution X-ray imaging devices~\cite{Parsafar:2015,Scott:2015}. CMOS technology can achieve pixel noise as low as a few \si{\electron} with pixel sizes as small as tens of \si{\square\um}~\cite{Bigas:2006jx,Park:2009ix,Seo:2015gz}.  Imaging areas of hundreds of \si{\square\cm} can be fabricated by standard processing of \SI{200}{\mm} (\SI{8}{\inch}) diameter silicon wafers at a rate of up to \num{1E5} per month by modern silicon foundries.  Amorphous selenium layers up to \SI{1}{\mm} thick are standard in the medical industry and would amount to up to 100\,\si{\gram} of \ce{^{82}Se} in a CMOS imager fabricated with isotopically enriched selenium.  Readout and digitization of the pixel signal can be performed on-board, simplifying the electronics as the data can be directly downloaded from each device. Ancillary infrastructure requirements can be minimal as the devices can be operated at room temperature. 

A large array of imagers can be deployed in an underground laboratory within a highly efficient cosmic ray veto~\cite{Agnes:2016fw} for the backgrounds in the energy region of interest (ROI) about \qbb\ to be dominated by radioactive contamination in the inner detector components.

The devices will resolve with high spatial resolution individual energy depositions. A two-dimensional projection of the ionization density of particle tracks occurring within exposures of a fraction of a second will be imaged on the plane of the pixel array.  The contributions in the ROI from $\alpha$'s and $\beta$'s emitted by the decay of uranium, thorium and their daughters in the bulk selenium or on the surfaces of the imagers will be suppressed to a negligible level by the established discrimination techniques to reject events from $\alpha$-decay by their distinct topology and sequences of radioactive decay by spatial correlations~\cite{AguilarArevalo:2015hf}.  The limiting background in the ROI will be electrons from interactions of long-range $\gamma$-rays emitted by the daughters \ce{^{214}Bi} and \ce{^{208}Tl} in the imagers, which cannot be effectively vetoed by time correlations with the primary short-range radiation due to the coarse time resolution of the devices. Other radioactive isotopes in the bulk selenium with $Q$-values above \qbb\ that may offer sub-dominant contributions to the background rate in the ROI are the long-lived cosmogenic isotope \ce{^{56}Co}, and in-situ cosmogenic activation of \ce{^{83}Se} and \ce{^{78,80,81,82}As}.

Other potential sources of $\gamma$-rays include the decays of \ce{^{214}Bi}, \ce{^{208}Tl} and cosmogenic isotopes in the detector components other than the imagers, as well as high-energy $\gamma$-rays emitted following neutron capture in the detector.  We expect to suppress these backgrounds by the appropriate selection and handling of construction materials to achieve activities of \ce{^{214}Bi}, \ce{^{208}Tl} and cosmogenic isotopes at the same level or lower than in the imagers. Background from $\gamma$-rays due to radiogenic neutrons from uranium and thorium decay can also be significantly reduced by the careful selection and arrangement of detector components for most neutrons to be captured without the emission of high-energy $\gamma$-rays.

\section{Background estimates}
\begin{figure}[t!]
\centering
\includegraphics[width=0.85\textwidth]{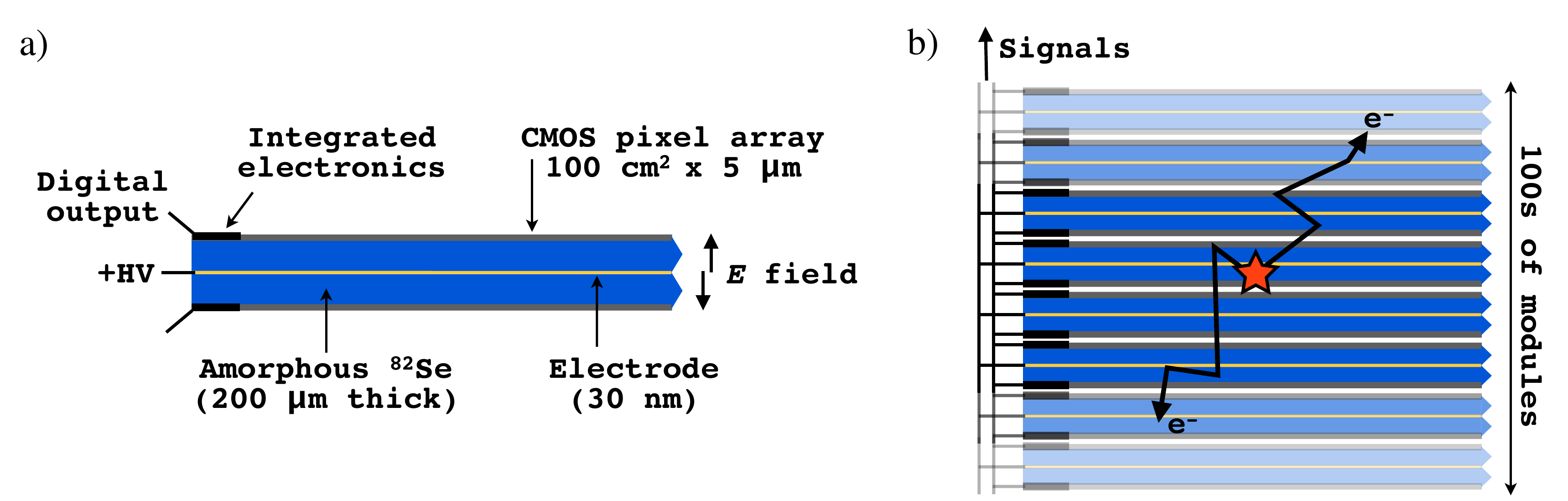}
\caption{Detector geometry implemented in the simulation: a) A detector module consisting of two back-to-back amorphous \ce{^{82}Se} high-resolution imagers that share a common high-voltage (HV) electrode. b) Segment of a tower of detector modules. The pattern repeats many times to build a tower of hundreds of modules. The two emitted electrons emerging from the double $\beta$ decay site (red star) are depicted.}
\label{fig:Sketch}
\end{figure}

We performed a Geant4~\cite{Agostinelli:2003fg,Allison:2006cd} simulation of a \ce{^{82}Se} CMOS imaging detector making realistic assumptions on the performance of the devices, with pixels \SI[product-units=power]{15 x 15}{\square\um} in size, a pixel readout noise of \SI{10}{\electron} and a pixel dynamic range of \SI{70}{\deci\bel}. We assumed a \SI{200}{\um}-thick active layer of amorphous \ce{^{82}Se} and a \SI{5}{\um}-thick dead layer on the front of each device corresponding to the pixel array structure. The detector consisted of towers of modules, each made of two back-to-back imagers that share a common electrode to which the high-voltage is applied. Figure~\ref{fig:Sketch} presents the simulated detector geometry.

To estimate the detector performance, we implemented full electron tracking and simulated the charge produced by ionization along the electron's path using a Fano model~\cite{Fano:1947cd}. Data on the intrinsic resolution of amorphous selenium is severely lacking, with measurements and Monte Carlo simulations existing only up to \SI{140}{\keV}~\cite{Blevis:1999ht, Fang:2012dp}. We adopted an average energy for the production of one electron-hole pair of \SI{15}{\electronvolt}, as measured for \SI{140}{\keV} X-rays at an electric field of \SI{30}{\volt\per\um}~\cite{Blevis:1999ht},  and a value of the Fano factor $F = 0.6$, as predicted at this field~\cite{Darbandi:2012bw}. From the simulation of the ionizing tracks and the instrumental response, followed by clustering and track reconstruction algorithms, we obtained a FWHM of the energy response at \qbb\ of \SI{24}{\keV}.

\begin{figure}[t!]
\centering
\includegraphics[width=0.6\textwidth]{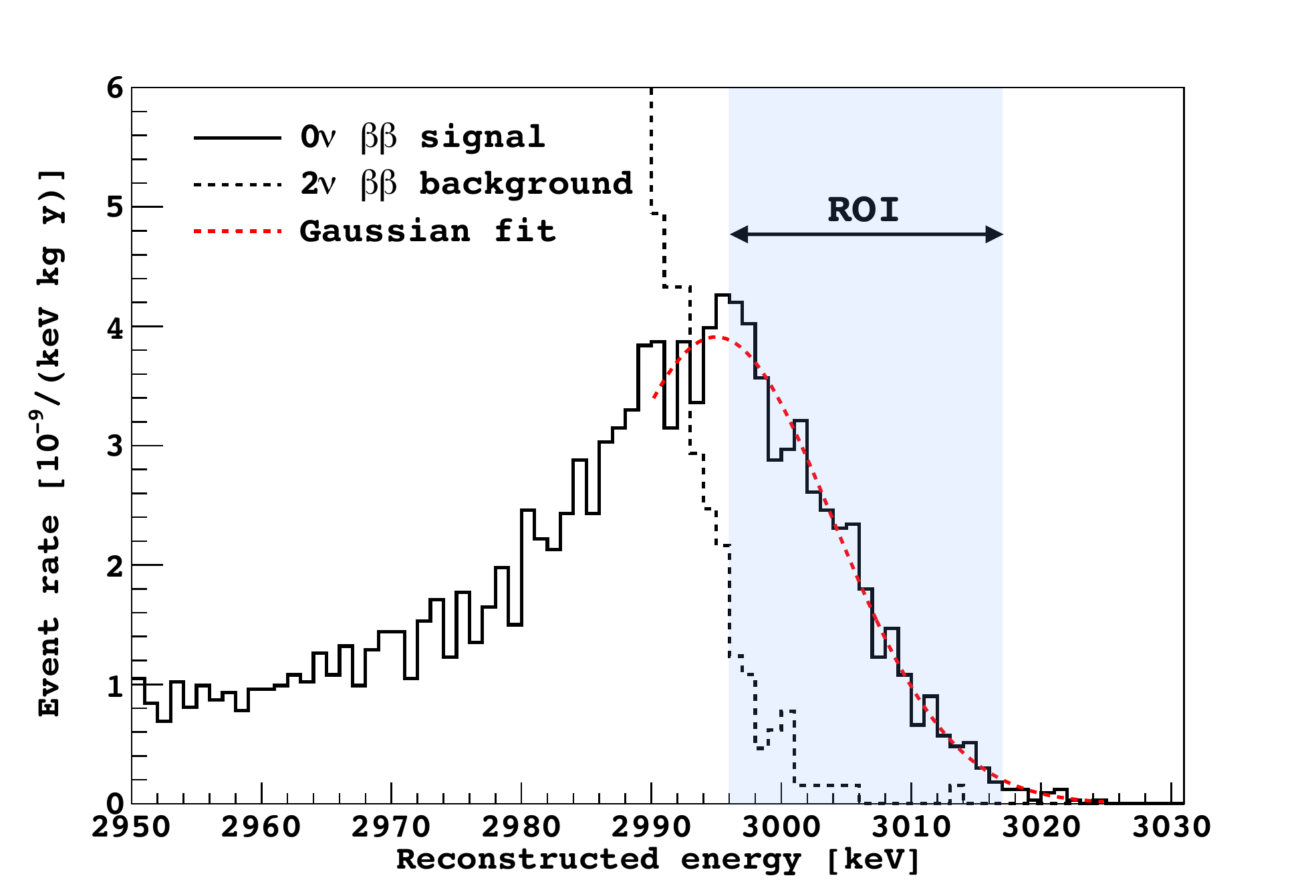}
\caption{Reconstructed energy spectra from simulated neutrinoless ($0\nu$) double $\beta$ decay with $\mbb = 1$\,\si{\meV\per\square\c} and a strongly quenched nucleon axial coupling constant, and two-neutrino ($2\nu$) double $\beta$ decay in the amorphous \ce{^{82}Se} imaging detector described in the text. The ROI for the neutrinoless double $\beta$ decay search is labeled.}
\label{fig:SeEightTwoSpectra}
\end{figure}

We first assessed the background from the two-neutrino decay by simulating pairs of electrons from \ce{^{82}Se} double $\beta$ decay. For two-neutrino decay, we assumed the measured decay rate of \SI{1.6}{\milli\becquerel\per\kg}~\cite{Arnold:2006bk}.  For the neutrinoless decay, we assumed the most unfavorable case, i.e., $\mbb = 1$\,\si{\meV\per\square\c} and a strongly quenched nucleon axial coupling constant, corresponding to a total signal rate of \SI{3.0E-7}{\per\kg\per\year}.  We simulated a common exposure of \SI{3E10}{\kg\year} to amass sufficient statistics for the spectra.  Initial momenta distributions for both decays were obtained with the DECAY0 program~\cite{Ponkratenko:2000im}.  The result is shown in Figure~\ref{fig:SeEightTwoSpectra}.  Even in this most unfavorable case, the neutrinoless decay would be distinguishable from background from the two-neutrino channel.  We defined a ROI extending from \SIrange{2996}{3017}{\keV}, selecting the start of the ROI slightly above the crossing point of the two spectra.  We determined that 15\% of the neutrinoless decays have a reconstructed energy in the ROI, corresponding to a signal rate of \SI{4.0E-8}{\per\kg\per\year}.  The two-neutrino background has a leakage in the ROI of \SI{5.1E-9}{\per\kg\per\year}, for a signal-to-background ratio of \num{8}.

Then we turned to the assessment of background from other sources.  We performed a full simulation of the expected electron backgrounds from radioactive decay assuming a bulk contamination in the inner detector of \ce{^{238}U}, \ce{^{232}Th} and their daughters of \SI{<110}{\micro\becquerel\per\kg}, as already achieved in enriched \ce{^{82}Se}~\cite{Beeman:2015du}. For the pixel array we assumed a \ce{^{238}U} and \ce{^{232}Th} contamination of \SI{<1E-2}{\micro\becquerel\per\square\cm}, and an additional \ce{^{210}Po} surface contamination of \SI{1E-1}{\micro\becquerel\per\square\cm}, as for the charge-coupled devices deployed in the DAMIC experiment~\cite{AguilarArevalo:2015hf}. We included the high-energy $\gamma$-rays produced by the capture of radiogenic neutrons from fission, $(\alpha,n)$ reactions of the \ce{^{238}U} and \ce{^{232}Th} decay chains, and $(\alpha,n)$ reactions of the out-of-equilibrium \ce{^{210}Po} activity.  We also considered the contribution in the bulk from cosmogenic \ce{^{56}Co} with an activity of \SI{<2E-2}{\micro\becquerel\per\kg}~\cite{Artusa:2016ek}, as well as activities of \ce{^{83}Se} and \ce{^{78,80,81,82}As} from in-situ cosmogenic production (at a depth of \num{3800} meters water-equivalent) of \SI{2E-2}{\micro\becquerel\per\kg} and \SI{3E-4}{\micro\becquerel\per\kg}, respectively~\cite{Bellini:2013kr, Albert:2016ba}. The first column in Table~\ref{tab:Background} summarizes the total expected rate in the ROI from these background sources.  For electrons emitted by $\beta$-decay we present the contribution from each class of events described above. For electrons produced by $\gamma$-ray interactions, we consider the total $\gamma$-ray flux from all background sources and present the contribution by the interaction mechanism.

\begin{table}[b!]
\centering
\caption{Estimated background rates of minimum-ionizing electron tracks produced by natural radioactivity in the amorphous \ce{^{82}Se} detector of \SI[product-units=power]{15 x 15}{\square\um} pixel size described in the text.  The first column specifies the origin of the background.  The second column gives the total rate observed in the ROI from the assumed radioactive contamination.  The third column presents the corresponding residual event rates after application of the discrimination criteria.
}
\label{tab:Background}
\smallskip
\begin{tabular}{|ccc|}
\hline
Radioactive background source		&Raw background rate				&Rate after discrimination \\
  							&[\si{\per\kg\per\year}]				&[\si{\per\kg\per\year}]\\
\hline
$\beta$-decay (bulk)		&\num{<3.3E-1}								&\num{<3.7E-9}\\
$\beta$-decay (surface)		&\num{<4.1E-1}							&\num{<1.2E-8}\\
$\beta$-decay (cosmogenic)		&\num{<9.9E-5}						&\num{<1.5E-7}\\
$\gamma$-ray (photoelectric)	&\num{<7.2E-4}							&\num{<7.2E-7}\\
$\gamma$-ray (Compton)		&\num{<1.6E-3}							&\num{<4.1E-7}\\
$\gamma$-ray (pair production)	&\num{<1.9E-6}							&\num{<1.9E-7}\\
Solar \ce{^{8}B} $\nu$		&~~~\num{3.2E-6}							&~~~\num{3.2E-9}\\
\hline
Total						&\num{<7.4E-1}							&\num{<1.5E-6}\\
\hline
\end{tabular}
\end{table}

The electron backgrounds from $\beta$-decay will be dominated by \ce{^{214}Bi} and \ce{^{208}Tl} in the bulk and on the surfaces of the imagers, as they are the only daughters from uranium and thorium that can emit electrons (the $\beta$~particle and, in some cases, conversion electrons) with total energy in the ROI. The exquisite spatial resolution and dead-time-free operation of the devices will allow to identify these decays within their corresponding decay sequences: \ce{^{218}Po}--\ce{^{214}Pb}--\ce{^{214}Bi}--\ce{^{214}Po} and \ce{^{216}Po}--\ce{^{212}Pb}--\ce{^{212}Bi}--\ce{^{208}Tl}. Considering a probability of missing a single decay in the sequence of \num{<1E-2}, a total suppression factor of \num{<1E-6} is expected. Likewise, the background rate from the dominant cosmogenic isotope, \ce{^{83}Se}, will be suppressed by a factor \num{<1E-2} by its spatial correlation with the subsequent $\beta$-decay of \ce{^{83}Br}.

\begin{sidewaysfigure}
\centering
\includegraphics[width=0.9\textwidth]{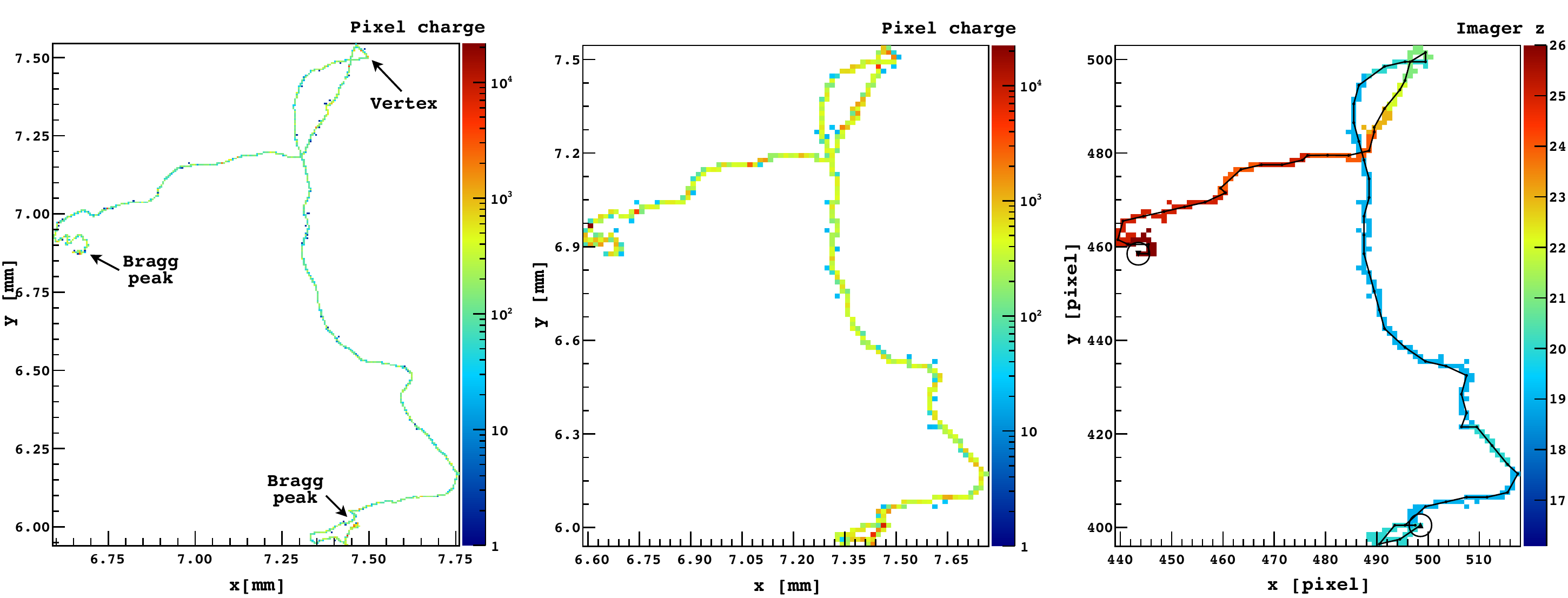}
\caption{Double electron event simulated with total energy in the neutrinoless decay ROI.  The electrons emerge from the vertex located at coordinates ($x$,$y$)~=~(\SI{7.5}{\mm}, \SI{7.5}{\mm}).  The color axis of the left and middle panels correspond to the number of charge carriers collected by a pixel. The left panel shows the simulated track with a \SI[product-units=power]{5 x 5}{\square\um} pixel size, while the middle and right panels present the simulated track with \SI[product-units=power]{15 x 15}{\square\um} technology. The color axis of the right panel corresponds to the identifier of the imager in the tower where the energy deposition took place; higher values represent imagers that are higher up in the tower. The black line shows the reconstructed track for the event, with circles marking where Bragg peaks were identified by the automated algorithm.}
\label{fig:DoubleTrackNoField}
\end{sidewaysfigure}

\begin{sidewaysfigure}
\centering
\includegraphics[width=0.9\textwidth]{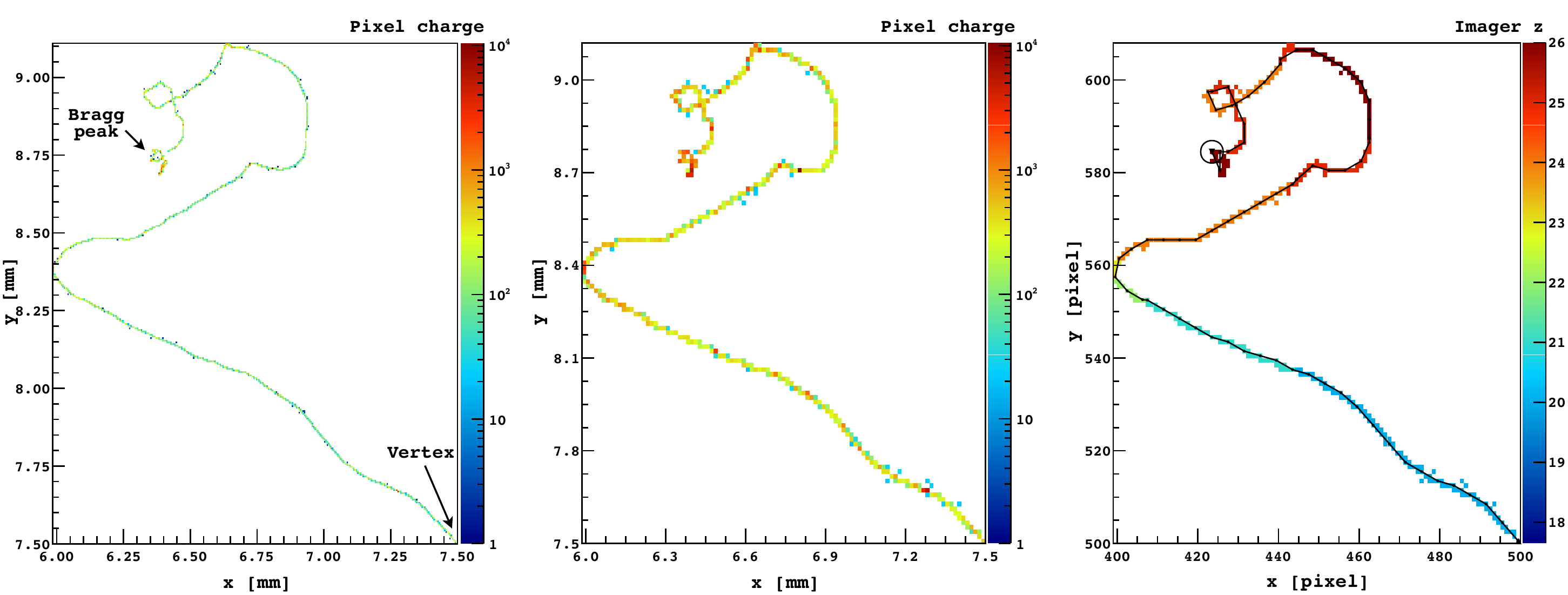}
\caption{Single electron event simulated with total energy in the neutrinoless decay ROI.  The electron emerges from the vertex located at coordinates ($x$,$y$)~=~(\SI{7.5}{\mm}, \SI{7.5}{\mm}).  The color axis of the left and middle panels correspond to the number of charge carriers collected by a pixel. The left panel shows the simulated track with a \SI[product-units=power]{5 x 5}{\square\um} pixel size, while the middle and right panels present the simulated track with \SI[product-units=power]{15 x 15}{\square\um} technology. The color axis of the right panel corresponds to the identifier of the imager in the tower where the energy deposition took place; higher values represent imagers that are higher up in the tower. The black line shows the reconstructed track for the event, with circles marking where Bragg peaks were identified by the automated algorithm.}
\label{fig:SingleTrackNoField}
\end{sidewaysfigure}

We have investigated the discrimination between double and single electron events by the analysis of their track topologies.  The proposed technology will reconstruct in extreme detail any ionizing track: Figures~\ref{fig:DoubleTrackNoField}~and~\ref{fig:SingleTrackNoField} show the reconstructed images of double electron and single electron events simulated in the ROI. The high quality of the images allows to separate, by simple visual examination, double electron tracks, with two Bragg peaks, from single electron tracks, with one Bragg peak.  We developed an algorithm based on the increase in straggling and stopping power of the electron at the Bragg peak to identify double electron events and suppress single electron background by \num{1E-3}. This technique is limited by single electrons that mimic double electron events due to the emission of a $\delta$-ray very close to the start of the track, where the emission vertex is indistinguishable from the electron starting point. We note that for a detector with an improved spatial resolution of \SI[product-units=power]{5 x 5}{\square\um} this limiting background reduces to \num{5E-4}.

We have also estimated the discrimination of events based on the spatial correlation with nearby energy depositions in the same exposure. The requirement that the event in the ROI is isolated, i.e., that there are no energy depositions within a \SI{5}{\cm} spherical region of the event, suppresses backgrounds from $\gamma$-ray pair production, Compton scattering of $\gamma$-rays with energies \SI{<3300}{\keV}, and $\beta$-decay of \ce{^{214}Bi} and \ce{^{208}Tl} with associated \SIrange{50}{300}{\keV} photon emission by a factor of \num{<1E-1}.

The third column of Table~\ref{tab:Background} presents the background for different classes of events after all selection criteria, approaching the ultimate requirement outlined in Eq.~\ref{eq:BackgroundRequirement}. The negligible background of single electrons from elastic scattering of \ce{^{8}B} solar $\nu$'s is presented for reference. The overall acceptance for double $\beta$ decay is 50\%, dominated by the inefficiency in the selection of events with two Bragg peaks.

We stress the modest requirements on the total background rate in the target necessary to achieve the final background levels presented in Table~\ref{tab:Background}. Furthermore, by not relying on self-shielding for the suppression of the external $\gamma$-ray flux, the final background levels are largely independent on the geometry and size of the detector array, and the proposed technology maximizes the fraction of the isotopically enriched target that can be effectively used for the search of neutrinoless double $\beta$ decay.

\begin{table}[b!]
\centering
\caption{ Signal and background rates in the neutrinoless double $\beta$ decay search assuming $\mbb = 1$\,\si{\meV\per\square\c} and different nucleon axial coupling constants. The second column gives the total rate of the neutrinoless decay, while the fourth and fifth columns give the final signal and background rates in the specified ROI after all discrimination criteria. We present both the case for a strongly quenched nucleon axial coupling constant ($g_{\rm phen.}$) and the case for a free nucleon axial coupling constant ($g_{\rm nucl.}$)~\cite{DellOro:2016gf}.
}
\label{tab:SB}
\smallskip
\begin{tabular}{|ccccc|}
\hline
Coupling constant		& Total signal rate 		& ROI &	Final signal rate 	& Final background rate \\
					& [\si{\per\kg\per\year}]	& [\si{\kilo\electronvolt}]	& [\si{\per\kg\per\year}]		&[\si{\per\kg\per\year}]\\
\hline
$g_{\rm phen.}$	&\num{3.0E-7}		&2996--3017	&\num{2.0E-8}				&\num{<1.5E-6} \\
$g_{\rm nucl.}$		&\num{7.1E-6}		& 2984--3005	&\num{8.2E-7}				&\num{<1.7E-6} \\
\hline
\end{tabular}
\end{table}

Table~\ref{tab:SB} compares the final background rate in the ROI with the neutrinoless decay rate under the assumption of $\mbb = 1$\,\si{\meV\per\square\c} and two different values for the nucleon axial coupling constant. To evaluate the final signal rate, we have applied the 50\% acceptance of the double-electron selection criteria to the initial neutrinoless decay rate in the specified ROI. For the canonical case, where the coupling is taken to be that of a free nucleon, the implementation of the proposed technology as described in this paper would already have the sensitivity to reach $\mbb = 1$\,\si{\meV\per\square\c} and probe the normal hierarchy of neutrino masses. A significant improvement in the signal-to-background is still necessary to achieve a comparable sensitivity to $\mbb$ for the case of a strongly quenched nucleon axial coupling constant.

\section{Further improvements}
A more precise reconstruction of the energy loss by electrons in the pixel array structure, and the optimization of the module geometry to minimize the fraction of dead material \textemdash\ by either increasing the thickness of the amorphous selenium layer or decreasing the thickness of the pixel array \textemdash\ could lead to an increase of the signal rate in the ROI by up to a factor of three without interfering background from the two-neutrino decay.

\begin{figure}[t!]
\centering
\includegraphics[width=0.62\textwidth]{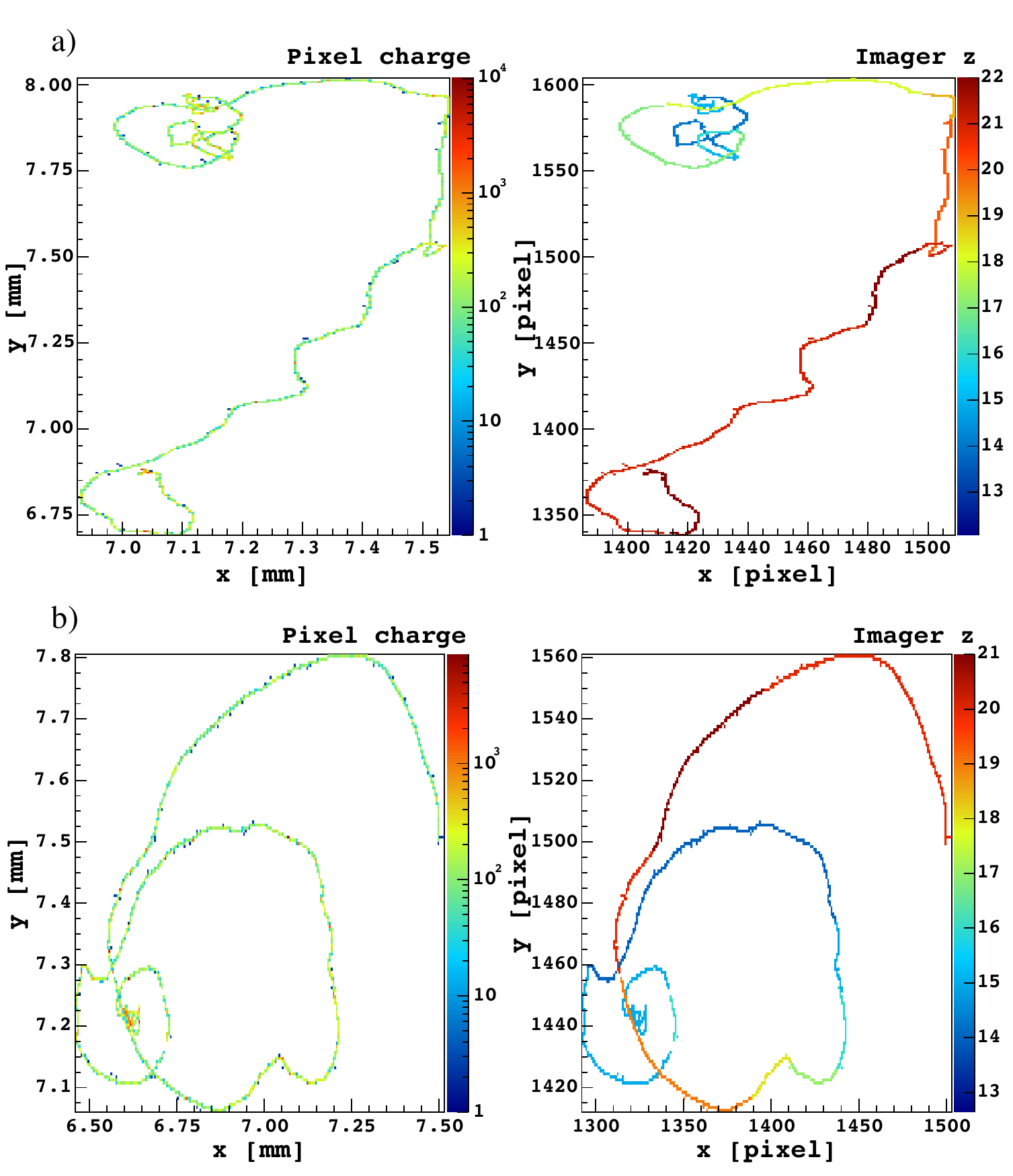}
\caption{Double electron (a) and single electron (b) events with vertex ($x$,$y$) = (\SI{7.5}{\mm}, \SI{7.5}{\mm}).  The events have been simulated assuming \SI[product-units=power]{5 x 5}{\square\um} pixel technology and a \SI{20}{\tesla} magnetic field perpendicular to the plane of the imagers.}
\label{fig:TracksWithField}
\end{figure}

Inspired by the mode of operation of bubble chambers~\cite{Glaser:1952gy,Hasert:1973dm,Hasert:1973hj,Hasert:1974eh}, we explored the possibility to operate the detector inside a magnet to enhance the topological signature of double $\beta$ decay events.  Figure~\ref{fig:TracksWithField} shows the tracks of double and single electron events in the ROI in the presence of a \SI{20}{\tesla} magnetic field: the change in the radius of curvature along the electron track provides a powerful and independent tool to discriminate between double and single electron events, which would improve the acceptance for the double electron signal and give an additional handle on potential backgrounds.  It would also provide valuable topological information on events from pair production by high-energy $\gamma$-rays.

Additionally, the detector's imaging capability will allow for the localization of the site of a double $\beta$ decay candidate within tens of \si{\ug}'s of selenium in a detector module.  Following the double $\beta$ decay of \ce{^{82}Se}, the \ce{^{82}Kr} daughter will remain essentially immobile~\cite{Greuter:1995ji}.  Its selective removal and detection by single atom counting~\cite{Chen:1999hh,Du:2003dx,Jiang:2012kd} may be possible and may offer an independent confirmation of double $\beta$ decay.

\section{Outlook}
We aim to establish an experimental program for the design, fabrication and operation of low noise CMOS imagers with amorphous selenium layers of natural isotopic abundance. Characterization of these devices with cosmic rays and radioactive sources in the laboratory will demonstrate the capability to image with high resolution individual ionizing tracks, and will allow to measure the intrinsic energy response of amorphous selenium to minimum ionizing particles, determining the average energy for the production of an electron-hole pair and the Fano factor.

A detector, consisting of an array of one hundred large-area imaging modules made from isotopically enriched \ce{^{82}Se}, operated in a shallow underground laboratory, will permit the experimental demonstration of the performance outlined in this paper. Ionizing tracks from two-neutrino double electron events near \qbb\ and single electrons produced by $\gamma$-ray calibration sources will allow to characterize the energy response and topological event discrimination. The uranium and thorium contamination of the imagers and the production cross-sections of cosmogenic isotopes in the \ce{^{82}Se} will also be directly measured. From these results it will be possible to extrapolate to the background expected in the ROI for a large detector operated in a low-radioactivity environment.

Considering the modular nature of the experiment, its compactness and simplicity in operation, the modest radio-purity requirements, and the scalability of the technology \textemdash\ which relies on widespread fabrication processes in the semiconductor industry \textemdash\ once the required performance is experimentally demonstrated, we would have set the path to a large detector capable of sweeping the parameter space of neutrinoless double $\beta$ decay in the inverted and normal hierarchies of neutrino masses.

\acknowledgments{
This work has been supported by the Kavli Institute for Cosmological Physics at the University of Chicago through grant NSF PHY-1125897 and an endowment from the Kavli Foundation, and by Princeton University through grant NSF PHY-1314507.  We thank R.~Bez, A.~Martini, P.~Organtini, and P.~Privitera for discussions and useful comments.}

\bibliographystyle{jhep}
\bibliography{ds}
\end{document}